\documentclass[12pt,reqno]{article}
\usepackage{amsfonts}
\usepackage{amsbsy}
\usepackage{graphicx}
\usepackage{amssymb,latexsym}
\usepackage{cite} 

\begin{document}

\title{Note on the matter coupling in Exotic General Massive 3D Gravity}
\begin{titlepage}
\author{Hakan Cebeci \footnote{E.mail:
hcebeci@eskisehir.edu.tr ; hcebeci@gmail.com}  \\
{\small Department of Physics, Eski\c{s}ehir Technical University, 26470 Eski\c{s}ehir, Turkey} }

\date{ } 

\maketitle

\bigskip

\begin{abstract}

\noindent 
In this work, within exterior algebra formalism, we revisit and study the matter coupling in Exotic General Massive 3D Gravity (EGMG), previously introduced in \cite{ozkan_1}. Considering that the matter Lagrangian is both metric co-frame and connection dependent in general, we explicitly construct associated source 2-form by algebraically solving auxiliary 1-form fields in terms of gravitational field variables, energy-momentum and hyper-momentum 2-forms. It is seen that that the source 2-form involves terms that are quadratic in energy-momentum and hyper-momentum 2-forms as well as terms that depend on covariant derivatives of them. In addition, we derive the consistency relation in order that matter-coupled EGMG field equation be third-way consistent. In general, the consistency requires a particular relation that should be satisfied by energy-momentum and hyper-momentum 2-forms. We note that our construction and formalism allows to investigate non-minimal matter couplings with gravitational fields. It also enables to construct the higher order curvature extensions of EGMG.

\vspace{1cm}


\end{abstract}

\end{titlepage}

\section{Introduction}

It is well-known that the general relativity in $ (2+1) $ dimensions with or without a cosmological constant has no local propagating degrees of freedom. However, it has been shown that Einstein theory of gravity with a negative cosmological constant admits asymptotically $ AdS_{3} $ black hole solutions and yields the massless gravitons \cite{banados}.   
The modification of the theory by adding gravitational Chern-Simons term and a negative cosmological constant has led to Topologically Massive Gravity (TMG) where the linearization of the field equations about $ AdS_{3} $ vacuum yields a  massive propagating graviton mode \cite{deser_1, deser_2}. This has further led to a research to investigate whether there exists a holographically dual conformal field theory (CFT) on a 2-dimensional boundary and thereby to construct a consistent quantum theory of gravity in three dimensions \cite{brown, witten}. On the other hand, the application of the methods developed in \cite{brown} has illustrated that TMG possesses some undesirable features in the sense that if the massive graviton mode has positive energy, one of the central charges calculated by Virasoro algebra turns out to be negative that leads to a non-unitary CFT on the boundary \cite{kraus}. In fact, to overcome the bulk-boundary clash problem and to set a consistent theory of gravity in the sense of unitarity, several massive gravity theories have been proposed. Note that TMG is a parity-violating third order massive gravity theory. The generalization of TMG has been achieved by introducing New Massive Gravity (NMG) which is a fourth-order parity preserving theory \cite{bergshoeff_1,bergshoeff_2}, where NMG involves parity-even quadratic curvature terms in the Lagrangian contrary to TMG that has parity-odd gravitational Chern-Simons term. Later, NMG has been further generalized to General Massive Gravity (GMG) where both parity-breaking gravitational Chern-Simons term and parity-preserving curvature terms are involved in the Lagrangian \cite{hohm}. Furthermore, an alternative theory to TMG has been proposed and identified as Minimal Massive Gravity (MMG) that involves a single propagating massive mode \cite{bergshoeff_3}. MMG is also a parity-violating massive gravity theory similar to TMG. However, unlike TMG, MMG is a third-way consistent theory where the consistency of MMG field equation can be achieved by re-using or iterating the MMG field equation not using the Bianchi identities. In fact, MMG is the first simplest third-way consistent massive gravity theory. Recently, a new massive gravity theory, dubbed Exotic Massive Gravity (EMG), has been introduced as a next-to simplest third-way consistent gravity theory \cite{ozkan_1}. This new theory is identified as exotic such that EMG Lagrangian involves parity odd terms. In fact, it has been obtained from a Chern-Simons-like parity-odd Lagrangian involving the dreibein, the spin connection and two auxiliary fields. We note that the action of EMG leads to a parity-preserving third-way consistent field equation. On the other hand, a consistent modification of EMG can be realized by introducing a cosmological parameter in the action. As is also mentioned in \cite{ozkan_1}, this further leads to a parity-violating generalization of EMG which can also be identified as Exotic General Massive Gravity (EGMG). 

It is also well-established that EMG and EGMG can be constructed by using the so-called Chern-Simons-like formulation. In the Chern-Simons-like formulation of 3d massive gravity theories, the massive gravity Lagrangian can include dreibein (or equivalently co-frame one-form), the spin connection and some auxiliary one-form fields \cite{afshar_1, bergshoeff_4, merbis }. In particular, in the so-called N flavour Chern-Simons-like formulation, massive gravity Lagrangian can be constructed by using $ (N - 2) $ auxiliary one-form fields in addition to dreibein and connection field. In this setting, for example MMG can be constructed by using $ N = 3 $ Chern-Simons-like formulation. On the other hand, NMG involves parity-even $ N = 4 $ formulation. Furthermore like NMG, EMG and EGMG can also be constructed by using an $ N = 4 $ Chern-Simons-like formulation. However, it should be noted that EMG involves a parity-preserving while EGMG possesses a parity-violating formulation. 
  
Note that both EMG and EGMG are higher-derivative extensions of TMG and therefore they are faced with unitarity problems. In particular, it has been illustrated that both EMG and EGMG involve propagating spin-2 modes that are ghost \cite{ozkan_1}. Furthermore, the central charges have been found to be negative which further implies the non-unitarity of holographically dual conformal field theory. To resolve the unitarity problem, bi-gravity models have been utilized. In fact, in \cite{ozkan_2}, a unitary extension of EMG has been constructed from a bi-gravity formulation. Although EMG involves unphysical ghost degrees of freedom, it should be noted that it admits some interesting vacuum solutions including BTZ and warped $ AdS_{3} $ black hole solutions as well as logarithmic deformations of $ AdS_{3} $ \cite{chernicoff, giribet_1}. Regarding the vacuum solutions of EMG, the conserved charges such as the mass and the angular momentum have been obtained in \cite{mann} and \cite{bergshoeff_5}. Furthermore, some other interesting physical features of EMG have been exhibited . For instance, in \cite{kilicarslan}, causality issue in EMG has been analyzed and a Birkhoff-like theorem in EMG has been presented. Again in \cite{bergshoeff_5}, the central charges in Chern-Simons like gravity theories including EMG have been obtained and an analysis of critical points such as chiral points has been performed in \cite{giribet_2}. On the other hand, in \cite{omar_rodriguez}, the canonical structure of EMG has been worked out by using Dirac Hamiltonian formalism.  

Another interesting issue that also deserves a particular investigation is the matter coupling in third-way consistent 3d massive gravity theories. It is known that unlike Topologically Massive Gravity (TMG) and New Massive Gravity (NMG) where the source tensor in matter-coupled TMG and NMG becomes the energy-momentum tensor itself, the matter coupling in third-way consistent 3d Massive Gravity theories including Minimal Massive Gravity (MMG) and Exotic Massive Gravity (EMG) requires a particular source tensor for the consistent coupling. In particular, in \cite{arvanitakis_2} where the minimal matter coupling in MMG has been first investigated, the source tensor has been constructed considering that the matter Lagrangian does not depend on the connection field. In addition, it has been shown that the source tensor depends on quadratic forms of energy-momentum tensor as well as its covariant derivatives. Later, in \cite{cebeci_1} where the minimal matter coupling in MMG has been examined within exterior algebra formalism, the source 2-forms have been constructed for the connection-independent matter coupling as well as the minimal spinor-matter coupling. In these works, it has also been illustrated that the source tensor (or source 2-form) should satisfy a consistency relation in order that MMG field equation be third-way consistent. On the other hand, in \cite{ozkan_1} where EMG and EGMG have been first introduced, the matter coupling has been investigated briefly considering that the matter Lagrangian depends on the dreibein field (corresponding to co-frame one-form field) only but not on the connection. In the related work, the source tensor of EMG field equation has been constructed by introducing and defining new Einstein-type and Schouten-type symmetric tensors. Furthermore, recently in \cite{deger_5}, an alternative method has been employed to reconstruct MMG and 3d massive gravity theories including EMG where in that work the matter coupling has also been investigated. In fact, in the related work, the authors obtain the field equations and derive the source tensor by defining new Einstein-like and Schouten-like tensors described in terms of dreibein and spin connection involving torsion. Interestingly, it has also been considered that the matter Lagrangian may involve both bosonic and fermionic matter couplings where the fermionic couplings can depend on the spin connection with torsion. In particular, in that work the authors have ultimately aimed and accomplished to construct the supersymmetric version of MMG which is the Minimal Massive Supergravity \cite{deger_4,deger_5}.    

In this work, employing exterior algebra notation, we revisit and perform a detailed investigation of the matter coupling in Exotic General Massive 3D Gravity (EGMG) considering that the matter Lagrangian is both co-frame and connection dependent in general but does not depend on auxiliary fields. In the present work, in order to construct source 2-form, we use exterior algebraic methods and obtain algebraic solutions to the auxiliary 1-form fields. To this end, we first derive the field equations from a variational principle by making independent variations of matter-coupled EGMG Lagrangian with respect to co-frame, the connection and two auxiliary fields. Then using the resulting field equations, we obtain algebraic solutions to the auxiliary 1-form fields in terms of Schouten 1-form, Cotton 2-form as well as the energy-momentum and the hyper-momentum 2-forms obtained from the co-frame and the connection variations of the matter Lagrangian respectively. It can be noted that our algebraic approach leads to the source 2-form that involves the terms that are quadratic in energy-momentum and hyper-momentum 2-forms as well as the terms that depend on covariant derivatives of them. In addition, we derive the consistency relation in order that matter-coupled EGMG field equation be third-way consistent. In general, the consistency requires a particular relation that should be satisfied by the energy-momentum and the hyper-momentum 2-forms. Note that our construction and formalism allows to examine both minimal and non-minimal matter couplings that may depend on the co-frame and the connection fields together with the matter field. We notice that our formalism also enables to construct higher order curvature extensions of EGMG.  

For the organization of the paper, in Section 2, we introduce the matter-coupled EGMG within exterior algebra formalism. where the matter Lagrangian is both co-frame and connection dependent in general. We derive the source 2-form for matter-coupled EGMG field equation by also investigating the third way consistency. In Section 3, to illustrate our formalism, we present some examples of minimal and non-minimal matter couplings. We further illustrate the higher curvature extensions of EGMG and finally we end up with some comments and conclusion.  

\section{ Matter coupling in Exotic General Massive 3D Gravity}
\label{section_1}
In exterior algebra formalism, the action of Exotic General Massive 3d Gravity (EGMG) coupled to matter can be expressed in the form
\begin{equation}
I = \int_{N} \, {\cal L} = \int_{N} ( {\cal L}_{EGMG} + {\cal L}_M ) 
\label{eqn_1_1}
\end{equation}
where the Lagrangian 3-form of EGMG can be written as
\begin{eqnarray}
& & {\cal L}_{EGMG} =  \frac{ \mu_{1} }{ 2 } \epsilon_{abc} \rho^{c} \wedge R^{ab}  + \frac{\mu_{2}}{6}  \epsilon_{abc} \rho^{a} \wedge \rho^{b} \wedge \rho^{c}  + \Lambda \ast 1    \nonumber \\
& & \frac{\mu_{3}}{2}  \rho_{c} \wedge D \rho^{c} + \mu_{4} \rho_{a} \wedge \ast e^{a} + \mu_{5} \lambda_{a} \wedge T^{a} \nonumber \\ 
& & + \frac{\mu_{6}}{2} \left( \omega^{a}\,_{b} \wedge d \omega^{b}\,_{a} + \frac{2}{3} \omega^{a}\,_{b} \wedge \omega^{b}\,_{c} \wedge \omega^{c}\,_{a} \right) \, 
\label{eqn_1_2}
\end{eqnarray}
and the 3-form $ {\cal L}_{M} = {\cal L}_{M} ( e , \omega , \phi_{j} ) $ denotes metric co-frame and connection dependent matter Lagrangian also involving collective matter (bosonic or fermionic) fields $ \phi_{j} $ where index $ j $ implies that the matter Lagrangian can involve more than one matter fields. Basic field variables involved in EGMG Lagrangian are metric co-frame 1-forms $ e^{a} $, in terms of which the spacetime metric can be decomposed as $ {\mbox{\boldmath ${g}$}}=\eta_{ab}\, e^{a} \otimes e^{b}$ with $\eta_{ab} = diag ( - + + )\,$ , connection 1-forms $ \omega^{a}\,_{b} $ and auxiliary 1-form fields $ \lambda^{a} $ and $ \rho^{a} $. $ \ast $ designates the Hodge operator which maps a $ p$-form to $ (3-p)$-form in 3 dimensions while $ D $ denotes the covariant exterior derivative with respect to connection $ \omega^{a}\,_{b} $ defined as
\begin{equation}
D \rho^{a} = d \rho^{a} + \omega^{a}\,_{b} \wedge \rho^{b} \, . 
\label{covariant_derivative}
\end{equation} 
Recall that $ e^{a} $ and $ \omega^{a}\,_{b} $ are gravitational field variables that satisfy the first Cartan structure equations 
\begin{equation}
d e^{a} + \omega^{a}\,_{b} \wedge  e^{b} = T^{a}
\label{eqn_1_3}
\end{equation}
where $T^{a}$ denotes torsion 2-forms while the
corresponding curvature 2-forms are obtained from the second Cartan
structure equations
\begin{equation}
R^{a b} = d \omega^{a b} + \omega^{a}\,_{c} \wedge \omega^{c\,b}\,\, . 
\label{eqn_1_4}
\end{equation}
We also consider that the spacetime is metric-compatible which implies that $\omega_{a \,b} = - \omega_{b \,a}\, $.
For future use, we also note that $R=\iota_{a} P^{a}$ is the curvature scalar that can be expressed in terms of Ricci 1-forms $P^{a} = \iota_{b} R^{ba} $ where $\iota_{b}$ is the inner product operator that obeys the identity $\iota_{b} e^{a} = \delta_{b}^{a} $. Auxiliary 1-form field $ \lambda_{a} $ can also be identified as the Lagrange multiplier 1-form that enforces the zero-torsion constraint for EGMG theory. 
$ \Lambda $ is the cosmological parameter while $ \mu_{i} $'s $ (i=1,\cdots 6) $ denote coupling constants. Also we note that for later convenience we impose the relation
\begin{equation}
\mu_{1} \mu_{2} =  \mu_{3}^{2} 
\label{coupling_constant}
\end{equation}
such that $ \mu_{3} \neq 0 $ as in \cite{ozkan_1}.\footnote{In the notation of \cite{ozkan_1}, the coupling constants read $ \mu_{1} = 1 $, $ \mu_{2} = \frac{ 1 }{ m^{4} } $, $ \mu_{3} = - \frac{ 1 }{ m^{ 2 } } $, $ \mu_{4}= \nu $, $ \mu_{5} = - m^{2} $, $ \mu_{6} = \frac{ 1 }{ 2 } ( \nu - m^{2} ) $ and hence the coupling constants obey the relation (\ref{coupling_constant}) as well. In our work, we take the coupling constants arbitrary except that there exists the relation (\ref{coupling_constant}) between them.} 
In fact, this relation would enable us to solve auxiliary one-form fields $ \lambda_{a} $ and $ \rho_{a} $ algebraically. Furthermore, it will be seen that EGMG is third-way consistent provided that the relation (\ref{coupling_constant}) holds between the coupling constants.\footnote{As a further remark, assigning dimensions $ [ e^{a} ] = [ L ] = [ M ]^{ - 1 } $ (with $ [e^{a}\,_{\mu} ] = [ 1 ] $), $ [ \omega^{ a }\,_{ b } ] = [ 1 ] $ (with $ [ \omega^{ a }\,_{ b \mu } ] = [ M ]$ ), $ [ \rho^{ a } ] = [ M ]^{2} $ and $ [ \lambda^{ a } ] = [ M ] $, it becomes that the cosmological parameter has mass dimension $ [ \Lambda ] = [ M ]^{5} $ and the coupling constants have mass dimensions $ [ \mu_{ 1 } ] = [ 1 ] $ (i.e $ \mu_{ 1 } $ is dimensionless), $ [ \mu_{2} ] = [ M ]^{-4} $, $ [ \mu_{3} ] = [ M ]^{-2} $, $ [ \mu_{4} ] = [ M ]^{2} $, $ [ \mu_{5} ] = [ M ]^{2} $ and $ [ \mu_{6} ] = [ M ]^{2} $. Notice that this choice of dimensions is related to parity even and parity odd assignment to auxiliary fields $ \rho^{a} $ and $ \lambda^{a} $. Also note that with this dimension assignment of field variables and coupling constants, the Lagrangian 3-form has dimensions of mass-squared. Then in order that the action be dimensionless, the Lagrangian should be multiplied by a constant $ \Lambda_{0} $ with mass dimension $ [ \Lambda_{0} ] = [ M ]^{-2} $. However this does not alter the field equations obtained from the variation of the action with respect to field variables. In addition, we note that by assigning alternative dimensions to auxiliary field variables conveniently, one can set all the coupling constants to unity.} 

Then, arbitrary independent variations of the matter-coupled EGMG action with respect to co-frames $e^{a}$, connection 1-forms $\omega^{a}\,_{b}$, the auxiliary fields $ \rho_{a} $ and $ \lambda_{a} $ results in the field equations   
\begin{equation}
\mu_{4} \epsilon_{abc} ( \rho^{a} \wedge e^{b} - \rho^{b} \wedge e^{a} )  + \Lambda \ast e_{c} + \mu_{5} D \lambda_{c}  +\tau_{c} = 0 \, ,
\label{eqn_1_5}
\end{equation}
\begin{equation}
\frac{ \mu_{1} }{ 2 } \epsilon^{abc} D \rho_{c}  - \frac{ \mu_{3} }{2} \rho^{a} \wedge \rho^{b} - \frac{\mu_{5}}{2} ( \lambda^{a} \wedge e^{b} - \lambda^{b} \wedge e^{a} ) + \mu_{6} R^{ba}  + \Omega^{ab} = 0 \, ,
\label{eqn_1_6}
\end{equation}
\begin{equation}
\frac{ \mu_{1} }{ 2 } \epsilon_{abc} R^{ab} + \frac{ \mu_{2} }{2} \epsilon_{abc} \rho^{a} \wedge \rho^{b} + \mu_{3} D \rho_{c} + \mu_{4} \ast e_{c} = 0 ,
\label{eqn_1_7}
\end{equation}
\begin{equation}
T^{a} = 0 \, 
\label{eqn_1_8}
\end{equation}
respectively. In addition, one can obtain the matter field equation by making independent variation with respect to matter fields $ \Phi_{j} $. For the case of a $p$-form bosonic matter field, matter field equation reads
\begin{equation}
\frac{ \delta {\cal L}_{M} }{ \delta \Phi_{j} } - (-1)^{p} d \left( \frac{ \delta {\cal L}_{M} }{\delta ( d \Phi_{j} )} \right) = 0 \, .
\label{eqn_1_9}
\end{equation} 
Recall that (\ref{eqn_1_5}) is derived from co-frame $ e^{a} $ variation where $ \tau_{c} $ denotes the energy-momentum 2-form obtained from the co-frame variation of matter Lagrangian and it is defined as
\begin{equation}
\tau_{c} = \frac{ \delta {\cal L}_{M}}{ \delta e^{c}} \, . 
\label{eqn_1_10}
\end{equation}
(\ref{eqn_1_6}) is obtained from connection $ \omega^{a}\,_{b} $ variation where 
\begin{equation}
\Omega^{a b} = \frac{ \delta {\cal L}_{M} }{ \delta \omega_{a b} }
\label{eqn_1_11}
\end{equation} 
is identified as hyper-momentum 2-form obtained from connection variation of matter Lagrangian. In addition, (\ref{eqn_1_7}) is obtained from auxiliary 1-form $ \rho^{a} $ variation while (\ref{eqn_1_8}) gives the torsion-free constraint obtained from auxiliary 1-form $ \lambda^{a} $ variation. Also note that the terms $ \rho^{a} \wedge e^{b} $ and $ \lambda^{a} \wedge e^{b} $ in the field equations (\ref{eqn_1_5}) and (\ref{eqn_1_6}) respectively have been anti-symmetrized. Now, to obtain algebraic solution for auxiliary 1-form field $ \lambda^{a} $, first we use equation (\ref{eqn_1_6}). In order to solve this equation for $ \lambda^{a} $, the term $ \epsilon^{abc} D \rho_{c} $ should be eliminated. We notice that this term can be eliminated by using equation (\ref{eqn_1_7}). If one further uses the relation (\ref{coupling_constant}), the equation (\ref{eqn_1_6}) takes the form
\begin{equation}
\lambda^{a} \wedge e^{b} - \lambda^{b} \wedge e^{a} = \frac{ 2 }{\mu_{5} } \left( \left( \frac{ \mu_{1}^{2} }{ 2 \mu_{3} } - \mu_{6} \right) R^{ab} - \frac{ \mu_{1} \mu_{4} }{ 2 \mu_{3} } \epsilon^{abc} \ast e_{c} + \Omega^{ab} \right) : = \Lambda^{ab} \, .
\label{eqn_1_12}
\end{equation}  
Referring to \cite{cebeci_1}, auxiliary one-form field $ \lambda^{a} $ can be obtained from  
\begin{equation}
\lambda^{a} = \iota_{b} \Lambda^{ba} - \frac{ 1 }{4} (\iota_{n} \iota_{p} \Lambda^{pn} ) e^{a} 
\label{lambda_1}
\end{equation}
where after some algebraic work, one can get 
\begin{equation}
\lambda^{a} = \frac{2}{\mu_{5} } \left( \left( \frac{\mu_{1}^{2}}{2 \mu_{3} } - \mu_{6} \right) Y^{a} + \frac{\mu_{1} \mu_{4} }{4 \mu_{3} } e^{a} + \xi^{a} \right) 
\label{lambda_2} 
\end{equation} 
where $ Y^{a} $ denotes the Schouten 1-forms defined as
\begin{equation}
Y^{a} = P^{a} - \frac{ 1 }{4} R e^{a}
\label{eqn_1_13}
\end{equation}
expressed in terms of Ricci 1-forms $ P^{a} $ and the curvature scalar $ R $. Also we define
\begin{equation}
\xi^{a} : = \iota_{b} \Omega^{ba} - \frac{ 1 }{4} (\iota_{n} \iota_{p} \Omega^{pn} ) e^{a} \, .
\label{eqn_1_14}
\end{equation}
Now to obtain an algebraic solution for $ \rho^{a} $, we use equation (\ref{eqn_1_5}). Then calculating $ D \lambda_{a} $ and substituting into (\ref{eqn_1_5}), the equation can be put into the form
\begin{eqnarray}
\rho^{a} \wedge e^{b} - \rho^{b} \wedge e^{a} & = & \frac{ 1 }{2 \mu_{4} } \left( 2 \left( \frac{\mu_{ 1 }^{ 2 } }{ 2 \mu_{3} } - \mu_{ 6 } \right) \epsilon^{abc} C_{c} + \Lambda \epsilon^{abc} \ast e_{c} \right. \nonumber \\
& & \left. + 2 \epsilon^{abc} D \xi_{c} + \epsilon^{abc} \tau_{c} \right) : = \bar{\Lambda}^{ab} 
\label{eqn_1_15}
\end{eqnarray}
where we identify $ C_{c} = D Y_{c} $ as Cotton 2-forms. Then similar to solution for $ \lambda^{a} $, the solution for $ \rho^{a} $ can be obtained from
\begin{equation}
\rho^{a} = \iota_{b} \bar{\Lambda}^{ba} - \frac{ 1 }{ 4 } ( \iota_{n} \iota_{p} \bar{\Lambda}^{pn} ) e^{a} 
\label{rho_1}
\end{equation}
where after some algebraic manipulations one can get
\begin{equation}
\rho^{a} = - \frac{ 1 }{ 2 \mu_{4} } \left( 2 \left( \frac{ \mu_{ 1 }^{ 2 } }{  2 \mu_{3} } - \mu_{ 6 } \right) \ast C^{a} + \frac{\Lambda}{2} e^{a} + 2 \chi^{a} + \bar{\tau}^{a} \right) 
\label{rho_2}
\end{equation}
where we have noted the identities 
$$ 
\epsilon^{abc} = \ast( e^{a} \wedge e^{b} \wedge e^{c} ) \, , \qquad  \iota_{c} C^{c} = 0 \, , \qquad \epsilon^{abc} \iota_{b} C_{c} = \ast C^{a} \, , \qquad \epsilon_{abc} \ast e^{ a } \wedge \ast C^{b} = 0 \, .
$$
In addition, we identify
\begin{equation}
\bar{\tau}^{a} : = \epsilon^{abc} \iota_{b} \tau_{c} - \frac{ 1 }{4} \epsilon^{nmc} ( \iota_{n} \iota_{m} \tau_{c} ) e^{a} 
\label{eqn_1_16}
\end{equation}
and define
\begin{equation}
\chi^{a} : = \epsilon^{abc} \iota_{b} D \xi_{c} - \frac{ 1 }{4} \epsilon^{nmc} ( \iota_{n} \iota_{m} D \xi_{c} ) e^{a} \, .
\label{eqn_1_17}
\end{equation} 
Note that writing the energy-momentum 2-forms $ \tau_{c} $ in the form $ \tau_{c} = \tau_{ca} \ast e ^{a} $, the expression (\ref{eqn_1_16}) can also be written as
\begin{equation}
\bar{\tau}^{a} = \frac{ 1 }{ 2 } \tau e^{a} - \tau_{n}\,^{a} e^{n} 
\label{eqn_1_18}
\end{equation}
where $ \tau $ can be identified as the trace of energy-momentum tensor that can be obtained from the relation $ \tau \ast 1 = e_{a} \wedge \tau^{a} $. In addition, by using the relations (\ref{eqn_1_16}) and (\ref{eqn_1_17}), one can derive the following identities : 
\begin{equation}
\epsilon_{abc} \ast e^{ a } \wedge \chi^{b} = \iota^{n} D \xi_{n} \wedge \ast e_{c} \, , \qquad \epsilon_{abc} \ast e^{ a } \wedge \bar{\tau}^{b} = \iota^{n} \tau_{n} \wedge \ast e_{c} \, . 
\label{tau_chi_identity}
\end{equation}
Then calculating $ \frac{ 1 }{2} \epsilon_{abc} \rho^{a} \wedge \rho^{b} $ and $ D \rho_{c} $ with $ \rho^{a} $ given in (\ref{rho_2}) and substituting into (\ref{eqn_1_7}), one can obtain the matter-coupled EGMG field equation in the form
\begin{equation} 
\frac{ 1 }{2} \bar{\mu}_{1} \epsilon_{abc} R^{ab} + \frac{ 1 }{2} \bar{\mu}_{2} \epsilon_{abc} \ast C^{a} \wedge \ast C^{b} + \bar{\mu}_{3} C_{c} + \bar{\mu}_{4} \ast e_{c} + \bar{\mu}_{5} D ( \ast C_{c} ) + \tilde{\tau}_{c} = 0
\label{eqn_1_19}     
\end{equation}
where the coefficients $ \bar{\mu}_{i} $ read
\begin{eqnarray}
& & \bar{\mu}_{1} = \mu_{1}  \, , \qquad  \bar{\mu}_{2} = \frac{ \mu_{2} }{ \mu_{4}^{2} } \left( \frac{ \mu_{ 1 }^{2} }{ 2 \mu_{3} }
- \mu_{ 6 } \right)^{2} \nonumber \\
& & \bar{\mu}_{3} = \frac{ \mu_{2} }{ 4 \mu_{4}^{2} } \left( \frac{ \mu_{ 1 }^{2} }{ 2 \mu_{3} } - \mu_{ 6 } \right) \Lambda \, , \qquad \bar{\mu}_{4} = \frac{ \mu_{2} \Lambda^{2} }{ 16 \mu_{4}^{2} }  + \mu_{4} \, , \nonumber \\
& & \bar{\mu}_{5} = - \frac{ \mu_{3} }{ \mu_{4} } \left( \frac{ \mu_{ 1 }^{2} }{ 2 \mu_{3} } - \mu_{ 6 } \right) \, .
\label{eqn_1_20}
\end{eqnarray}
Here $ \tilde{\tau}_{c} $ can be identified as the source 2-forms of matter-coupled EGMG field equation. Further, it can be noted that the following identities that read 
\begin{eqnarray} 
\epsilon_{abc} \ast C^{a} \wedge e^{b} = C_{c} \, , \qquad \epsilon_{abc} e^{a} \wedge \bar{\tau}^{b} = \tau_{c} \, , \qquad \epsilon_{abc} e^{a} \wedge \chi^{b} = D \xi_{c} \, , 
\label{eqn_1_21}
\end{eqnarray} 
hold. Then we derive the source 2-forms $ \tilde{\tau}_{c} $ as 
\begin{eqnarray}
\tilde{\tau}_{c} & = & \frac{ \mu_{2} \Lambda }{ 8 \mu_{4}^{2} } \tau_{c} - \frac{ \mu_{3} }{ 2 \mu_{4} } D \bar{\tau}_{c} - \frac{ \mu_{3 }}{ \mu_{4} } D \chi_{c} + \frac{ \mu_{2} }{ \mu_{4}^{2} } \left( \frac{ \mu_{ 1 }^{2} }{ 2 \mu_{3} } - \mu_{ 6 } \right) \epsilon_{abc} \ast C^{a} \wedge \chi^{b} \nonumber \\
& + & \frac{ \mu_{2} }{ 2 \mu_{4}^{2} } \left( \frac{ \mu_{ 1 }^{2} }{ 2 \mu_{3} } - \mu_{ 6 } \right) \epsilon_{abc} \ast C^{a} \wedge \bar{ \tau }^{b} + \frac{ \mu_{2} \Lambda }{ 4 \mu_{4}^{2} } D \xi_{c} \nonumber \\ 
& + & \frac{ \mu_{2} }{ 2 \mu_{4}^{2} } \epsilon_{abc} \chi^{a} \wedge \chi^{b} + \frac{ \mu_{2} }{ 2 \mu_{4}^{2} } \epsilon_{abc} \bar{\tau}^{a} \wedge \chi^{b} + \frac{ \mu_{2} }{ 8 \mu_{4}^{2} } \epsilon_{abc} \bar{\tau}^{a} \wedge \bar{\tau}^{b} \,.
\label{source_term}  
\end{eqnarray} 
It can be seen that when $ \Lambda \neq 0 $, the source term of EGMG field equation involves a leading term that is linear in energy-momentum 2-form $ \tau_{c} $. It also involves the terms that are quadratic in energy-momentum and hyper-momentum 2-forms as well as terms that depend on covariant derivatives of them. Also note that when the matter Lagrangian is connection-independent (i.e $ \Omega^{ab} = 0 $), the source 2-form $ \bar{\tau}_{c} $, when expressed in the form $ \bar{\tau}_{c} = \bar{\tau}_{cb} \ast e^{b} $, the source tensor $ \bar{\tau}_{cb} $ reduces to the constructions obtained in \cite{ozkan_1} and \cite{deger_5}. Next, we examine the third-way consistency of matter-coupled EGMG field equation (\ref{eqn_1_19}). To this end, we act covariant exterior derivative operator $ D $ on the field equation. First, notice that Einstein 2-forms $ G_{c} $ defined by
\begin{equation}
G_{c} = - \frac{ 1 }{2} R^{ab} \wedge \ast ( e_{a} \wedge e_{b} \wedge e_{c} ) =  - \frac{ 1 }{2} \epsilon_{abc} R^{ab}\
\label{eqn_1_22}
\end{equation} 
satisfy the identity (for a metric-compatible spacetime) $ D G_{c} = 0 $. In addition, the Cotton 2-forms $ C_{c} $ satisfy the relation (that holds in torsion-free spacetimes) 
\begin{equation}
D C_{c} =  D^{2} Y_{c} = R_{cb} \wedge Y^{b} = 0 
\label{eqn_1_23}
\end{equation}
where we have used the curvature-Schouten relation
\begin{equation}
R_{cb} = Y_{c} \wedge e_{b} - Y_{b} \wedge e_{c}  
\label{curvature_shouten}
\end{equation}
and noted that $ e_{b} \wedge Y^{b} = 0 $ (for $ T^{a} = 0 $). Also, we point out that $ D ( \ast e_{c} ) = 0 $ for a torsion-free spacetime. Then acting operator $ D $ on matter-coupled EGMG equation and noting that 
\begin{equation}
D^{2} \ast C_{c} = R_{cb} \wedge \ast C^{b} \, , 
\label{eqn_1_24}
\end{equation}  
one obtains
\begin{equation}
\bar{\mu}_{2} \epsilon_{abc} D ( \ast C^{a} ) \wedge \ast C^{b} + \bar{\mu}_{5} R_{cb} \wedge \ast C^{b} + D \tilde{\tau}_{c} = 0 \, .
\label{eqn_1_25}
\end{equation}
By using (\ref{eqn_1_19}) one can eliminate the term $ D ( \ast C^{a} ) $ for third-way consistency. In addition, using the relation (\ref{coupling_constant}) between coupling constants, one can derive the consistency relation in the form
\begin{equation}
D \tilde{\tau}_{c} = - \frac{ \mu_{2} }{ \mu_{4} \mu_{3} } \left( \frac{ \mu_{ 1 }^{2} }{ 2 \mu_{3} } - \mu_{ 6 } \right) \epsilon_{abc} \tilde{\tau}^{a} \wedge \ast C^{b} \, .
\label{consistency}
\end{equation}
Then the consistency of matter-coupled EGMG field equation (\ref{eqn_1_19}) requires the source term $ \tilde{\tau}_{c} $ presented in (\ref{source_term}) to satisfy the relation (\ref{consistency}). Now let us calculate the left and the right hand sides of the consistency relation (\ref{consistency}) explicitly in order to see that for the source 2-form given in (\ref{source_term}) whether both sides are identically equal or an additional relation should be imposed on the energy-momentum and hyper-momentum 2-forms in order that the matter-coupled field equation be consistent. Then using the relations
\begin{equation}
D^{2} \bar{\tau}_{c} = R_{cb} \wedge \bar{\tau}^{b}  \, , \quad D^{2} \chi_{c} = R_{cb} \wedge \chi^{b} \, , \quad D^{2} \xi_{c} = R_{cb} \wedge \xi^{b} 
\label{eqn_1_26}
\end{equation}
the left hand side of the consistency relation can be calculated as
\begin{eqnarray}
D \tilde{\tau}_{c} & = & \frac{ \mu_{2} \Lambda }{ 8 \mu_{4}^{2} } D \tau_{c} - \frac{ \mu_{3} }{ 2 \mu_{4} } R_{cb} \wedge \bar{\tau}^{b} - \frac{ \mu_{3} }{ \mu_{4} } R_{cb} \wedge \chi^{b} \nonumber \\ 
& + &  \frac{ \mu_{2} }{ \mu_{4}^{2} } \left( \frac{ \mu_{1}^{2} }{ 2 \mu_{3} } - \mu_{6} \right) \epsilon_{abc} D ( \ast C^{a} ) \wedge \chi^{b} \nonumber \\
& - & \frac{ \mu_{2} }{ \mu_{4}^{2} } \left( \frac{ \mu_{1}^{2} }{ 2 \mu_{3} } - \mu_{6} \right) \epsilon_{abc} \ast C^{a} \wedge D \chi^{b} \nonumber \\
& + & \frac{ \mu_{2} }{ 2 \mu_{4}^{2} } \left( \frac{ \mu_{1}^{2} }{ 2 \mu_{3} } - \mu_{6} \right) \epsilon_{abc} D ( \ast C^{a} ) \wedge \bar{\tau}^{b} \nonumber \\
& - & \frac{ \mu_{2} }{ 2 \mu_{4}^{2} } \left( \frac{ \mu_{1}^{2} }{ 2 \mu_{3} } - \mu_{6} \right) \epsilon_{abc} \ast C^{a} \wedge D \bar{\tau}^{b} + \frac{ \mu_{2} \Lambda }{ 4 \mu_{4}^{2} } R_{cb} \wedge \xi^{b} \nonumber \\ 
& + & \frac{ \mu_{2} }{ \mu_{4}^{2} } \, \epsilon_{abc} \, D \chi^{a} \wedge \chi^{b}  
+ \frac{ \mu_{2} }{ 2 \mu_{4}^{2} } \, \epsilon_{abc} \, D \bar{\tau}^{a} \wedge \chi^{b} \nonumber \\
& - & \frac{ \mu_{2} }{ 2 \mu_{4}^{2} } \, \epsilon_{abc} \, \bar{\tau}^{a} \wedge D \chi^{b} 
+ \frac{ \mu_{2} }{ 4 \mu_{4}^{2} } \, \epsilon_{abc} \, D \bar{\tau}^{a} \wedge \bar{\tau}^{b} \, .
\label{eqn_1_27}
\end{eqnarray}
Next, we eliminate the term $ D ( \ast C^{a} ) $ from the field equation (\ref{eqn_1_19}) and again use the relation (\ref{coupling_constant}). After lengthy calculations and simplifications, the left hand side of the consistency relation can be calculated as
\begin{eqnarray}
LHS : = D \tilde{\tau}_{c} & = & \frac{ \mu_{2} \, \Lambda }{ 8 \mu_{4}^{2} } D \tau_{c} + \frac{ \mu_{2} \bar{\mu}_{2} }{ \mu_{3} \mu_{4} } \ast C_{c} \wedge \ast C_{b} \wedge \chi^{b} + \frac{ \mu_{2} \bar{\mu}_{3} }{\mu_{3} \mu_{4} } \epsilon_{abc} \, C^{a} \wedge \chi^{b} \nonumber \\  
& + & \frac{ \mu_{2} \bar{\mu}_{4} }{ \mu_{3} \, \mu_{4} } \iota_{n} D \xi^{n} \wedge \ast e_{c} + \frac{ \mu_{2}^{2} \,\Lambda }{ 8 \mu_{3} \mu_{4}^{3} } \epsilon_{abc} \, \tau^{ a } \wedge \chi^{b}  \nonumber \\ 
& - & \frac{ \mu_{2}^{2} }{ \mu_{3} \, \mu_{4}^{3} } \left( \frac{ \mu_{ 1 }^{2} }{ 2 \mu_{3} } - \mu_{ 6 } \right) \ast C_{b} \wedge \chi_{c} \wedge \chi^{b} \nonumber \\
& - & \frac{ \mu_{2}^{2} }{ 2 \mu_{3} \, \mu_{4}^{3} } \left( \frac{ \mu_{ 1 }^{2} }{ 2 \mu_{3} } - \mu_{ 6 } \right) \ast C_{b} \wedge \bar{\tau}_{c} \wedge \chi^{b} + \frac{ \mu_{2}^{2} \, \Lambda }{ 4 \mu_{3} \mu_{4}^{3} } \epsilon_{abc} D \xi^{a} \wedge \chi^{b} \nonumber \\ 
& - & \frac{ \mu_{2} }{ \mu_{4}^{2} } \left( \frac{ \mu_{ 1 }^{2} }{ 2 \mu_{3} } - \mu_{ 6 } \right) \epsilon_{abc} \ast C^{a} \wedge D \chi^{b}  + \frac{ \mu_{2} \bar{\mu}_{3} }{ 2 \mu_{3}  \mu_{4} } \epsilon_{abc} C^{a} \wedge \bar{\tau}^{b} \nonumber \\
& +  & \frac{ \mu_{2} \bar{\mu}_{2} }{ 2 \mu_{3} \, \mu_{4} } \ast C_{c} \wedge \ast C_{b} \wedge \bar{\tau}^{b} + \frac{ \mu_{2} \bar{\mu}_{4} }{ 2 \mu_{3} \, \mu_{4} } \iota_{n} \tau^{n} \wedge \ast e_{c} \nonumber \\
& + & \frac{ \mu_{2}^{2} \, \Lambda }{ 16 \mu_{3} \, \mu_{4}^{3} } \epsilon_{abc} \tau^{a} \wedge \bar{\tau}^{b} 
- \frac{ \mu_{2}^{2} }{ 2 \mu_{3} \mu_{4}^{3} } \left( \frac{ \mu_{ 1 }^{2} }{ 2 \mu_{3} } - \mu_{ 6 } \right) \ast C_{b} \wedge \chi_{c} \wedge \bar{\tau}^{b} \nonumber \\
& - & \frac{ \mu_{2}^{2} }{ 4 \mu_{3} \, \mu_{4}^{3} } \left( \frac{ \mu_{ 1 }^{ 2 } }{ 2 \mu_{3} } - \mu_{ 6 } \right) \ast C_{b} \wedge \bar{\tau}_{c} \wedge \bar{\tau}^{b}  + \frac{ \mu_{2}^{2} \, \Lambda }{ 8 \mu_{3} \mu_{4}^{3} } \epsilon_{abc} D \xi^{a} \wedge \bar{\tau}^{b} \nonumber \\
& - & \frac{ \mu_{2} }{ 2 \mu_{4}^{2} } \left( \frac{ \mu_{ 1 }^{ 2 } }{ 2 \mu_{3} } - \mu_{ 6 } \right) \epsilon_{abc}  \ast C^{a} \wedge D \bar{\tau}^{b} + \frac{ \mu_{2}  \Lambda }{ 4 \mu_{4}^{2} } R_{cb} \wedge \xi^{b} \, . 
\label{eqn_1_28}
\end{eqnarray}
Next by using the expression (\ref{source_term}) for the source 2-form $\tilde{\tau}^{a}$, the right hand side of the consistency relation can be obtained in the form
\begin{eqnarray}
RHS & : = & - \frac{ \mu_{2} }{ \mu_{4} \mu_{3} } \left( \frac{ \mu_{ 1 }^{ 2 } }{ 2 \mu_{3} } - \mu_{ 6 } \right) \epsilon_{abc} \, \tilde{\tau}^{a} \wedge \ast C^{b} \nonumber \\
& = & - \frac{ \mu_{2}^{2} \, \Lambda }{ 8 \mu_{3} \, \mu_{4}^{3} } \left( \frac{ \mu_{ 1 }^{ 2 } }{ 2 \mu_{3} } - \mu_{ 6 } \right) \epsilon_{abc} \, \tau^{a} \wedge \ast C^{b} + \frac{ \mu_{2} }{ 2 \mu_{4}^{2} } \left( \frac{ \mu_{ 1 }^{ 2 } }{ 2 \mu_{3} } - \mu_{ 6 } \right) \epsilon_{abc} \, D \bar{\tau}^{a} \wedge \ast C^{b} \nonumber \\
& + & \frac{ \mu_{2} }{ \mu_{4}^{2} } \left( \frac{ \mu_{ 1 }^{ 2 } }{ 2 \mu_{3} } - \mu_{ 6 } \right) \epsilon_{abc} \, D \chi^{a} \wedge \ast C^{b} - \frac{ \mu_{2}^{2} }{ \mu_{3} \mu_{4}^{3} } \left( \frac{ \mu_{ 1 }^{ 2 } }{ 2 \mu_{3} } - \mu_{ 6 } \right)^{2} \ast C_{c} \wedge \chi_{b} \wedge \ast C^{b} \nonumber \\
& - & \frac{ \mu_{2}^{2} }{ 2 \mu_{3} \mu_{4}^{3} } \left( \frac{ \mu_{ 1 }^{ 2 } }{ 2 \mu_{3} } - \mu_{ 6 } \right)^{2} \ast C_{c} \wedge \bar{\tau}_{b} \wedge \ast C^{b} - \frac{ \mu_{2}^{2} \, \Lambda }{ 4 \mu_{3} \mu_{4}^{3} } \left( \frac{ \mu_{ 1 }^{ 2 } }{ 2 \mu_{3} } - \mu_{ 6 } \right) \epsilon_{abc} \, D \xi^{a} \wedge \ast C^{b} \nonumber \\
& - & \frac{ \mu_{2}^{2} }{ \mu_{3} \mu_{4}^{3} }  \left( \frac{ \mu_{ 1 }^{ 2 } }{ 2 \mu_{3} } - \mu_{ 6 } \right) \chi_{c} \wedge \chi_{b} \wedge \ast C^{b} \nonumber \\
& + & \frac{ \mu_{2}^{2} }{ 2 \mu_{3} \mu_{4}^{3} } \left( \frac{ \mu_{ 1 }^{ 2 } }{ 2 \mu_{3} } - \mu_{ 6 }  \right) \left( \bar{\tau}_{b} \wedge \chi_{c} \wedge \ast C^{b} - \bar{\tau}_{c} \wedge \chi_{b} \wedge \ast C^{b} \right) \nonumber \\
& - & \frac{ \mu_{2}^{2} }{ 4 \mu_{3} \mu_{4}^{3} } \left( \frac{ \mu_{ 1 }^{ 2 } }{ 2 \mu_{3} } - \mu_{ 6 } \right) \bar{\tau}_{c} \wedge \bar{\tau}_{b} \wedge \ast C^{b} \, . 
\label{eqn_1_29}
\end{eqnarray}
Then we compare the left and the right hand sides of the consistency relation (\ref{consistency}). It can be seen that the coefficients of the terms $ \epsilon_{abc} D \bar{\tau}^{a} \wedge \ast C^{b} $, $ \epsilon_{abc} D \chi^{a} \wedge \ast C^{b} $, $ \ast C_{c} \wedge \chi_{b} \wedge \ast C^{b} $, $ \ast C_{c} \wedge \bar{\tau}_{b} \wedge \ast C^{b} $, $ \chi_{c} \wedge \chi_{b} \wedge \ast C^{b} $, $ \bar{\tau}_{b} \wedge \chi_{c} \wedge \ast C^{b} $, $ \bar{\tau}_{c} \wedge \chi_{b} \wedge \ast C^{b} $ and $ \bar{\tau}_{c} \wedge \bar{\tau}_{b} \wedge \ast C^{b} $ on the left and the right hand sides become equal, hence these terms will cancel. Then in order that the matter-coupled EGMG field equation (\ref{eqn_1_19}) be third-way consistent, the consistency requires that the additional relation should be imposed between the energy-momentum and hyper-momentum 2-forms. It explicitly reads 
\begin{eqnarray}
& & \frac{ \mu_{2} \, \Lambda }{ 8 \mu_{4}^{2} } D \tau_{c} + \frac{ \mu_{2} \bar{\mu}_{3} }{ \mu_{3} \mu_{4} } \epsilon_{abc} C^{a} \wedge \chi^{b} + \frac{ \mu_{2} \bar{\mu}_{4} }{ \mu_{3} \mu_{4} } \iota_{n} D \xi^{n} \wedge \ast e_{c} + \frac{ \mu_{2}^{2} \, \Lambda }{ 8 \mu_{3} \, \mu_{4}^{3} } \epsilon_{abc} \tau^{a} \wedge \chi^{b} \nonumber \\
& & + \frac{ \mu_{2}^{2} \, \Lambda }{ 4 \mu_{3} \, \mu_{4}^{3} } \epsilon_{abc} D \xi^{a} \wedge \chi^{b} + \frac{ \mu_{2} \bar{\mu}_{3} }{ 2 \mu_{3} \mu_{4} } \epsilon_{abc} C^{a} \wedge \bar{\tau}^{b} + \frac{ \mu_{2} \bar{\mu}_{4} }{ 2 \mu_{3} \mu_{4} } \iota_{n} \tau^{n} \wedge \ast e_{c}  \nonumber \\
& & + \frac{ \mu_{2}^{2} \, \Lambda }{ 16 \mu_{3} \, \mu_{4}^{3} } \epsilon_{abc} \tau^{a} \wedge \bar{\tau}^{b}  + \frac{ \mu_{2}^{2} \, \Lambda }{ 8 \mu_{3} \, \mu_{4}^{3} } \epsilon_{abc} D \xi^{a} \wedge \bar{\tau}^{b}  + \frac{ \mu_{2} \, \Lambda }{ 4 \mu_{4}^{2} } R_{cb} \wedge \xi^{b} \nonumber \\
& & =  - \frac{ \mu_{2}^{2} \, \Lambda }{ 4 \mu_{3} \, \mu_{4}^{3} } \left( \frac{ 1 }{2} \epsilon_{abc} \tau^{a} \wedge \ast C^{b} + \epsilon_{abc} D \xi^{a} \wedge \ast C^{b} \right) \, .
\label{eqn_1_30} 
\end{eqnarray} 
The relation (\ref{eqn_1_30}) can also be identified as the consistency condition. From the relation, it can be seen that for the case where $ \Lambda = 0 $, the right hand side of the expression (\ref{eqn_1_30}) vanishes. For the left hand side, it becomes that the coefficients $ \bar{\mu}_{3} = 0 $ and 
$ \bar{\mu}_{4} = \mu_{4} $. Then for the case of vanishing cosmological parameter $ \Lambda $, the additional relation reduces to a simple expression of the form
\begin{equation}
\iota_{n} D \xi^{n} \wedge \ast e_{c} + \frac{ 1 }{ 2 } \iota_{n} \tau^{n} \wedge \ast e_{c} = 0 \, .
\label{eqn_1_31}
\end{equation}    
On the other hand, if matter Lagrangian is independent of the connection 1-form 
$ \omega^{a}\,_{b} $ but just depends on the co-frame 1-form $ e^{a} $ and the matter fields, then we have $ \Omega^{ab} = 0 $ i.e the hyper-momentum forms vanish. In addition, it becomes that $ \xi^{a} = 0 $ and $ \chi^{a} = 0 $. Then the expression (\ref{eqn_1_30}) reduces to (for $ \Lambda \neq 0 $)
\begin{eqnarray}
& & \frac{ \mu_{2} \, \Lambda }{ 8 \mu_{4}^{2} } D \tau_{c} + \frac{ \mu_{2} \bar{\mu}_{3} }{ 2 \mu_{3} \mu_{4} } \epsilon_{abc} \, C^{a} \wedge \bar{\tau}^{b} + \frac{ \mu_{2} \bar{\mu}_{4} }{ 2 \mu_{3} \mu_{4} } \iota_{n} \tau^{n} \wedge \ast e_{c} + \frac{ \mu_{2}^{2} \, \Lambda }{ 16 \mu_{3} \, \mu_{4}^{3} } \epsilon_{abc} \, \tau^{a} \wedge \bar{\tau}^{b} \nonumber \\
& & = - \frac{ \mu_{2}^{2} \, \Lambda }{ 8 \mu_{3}  \, \mu_{4}^{3} } \left( \frac{ \mu_{ 1 }^{ 2 } }{ 2 \mu_{3} } - \mu_{ 6 } \right) \epsilon_{abc} \, \tau^{a} \wedge \ast C^{b} \, .   
\label{eqn_1_32}
\end{eqnarray} 
To check the consistency, let us compare the left and the right hand sides of the expression (\ref{eqn_1_32}). 
Note that if the matter Lagrangian is independent of the connection and involves bosonic (minimal) matter coupling only, we have $ D \tau_{c} = 0 $ i.e the energy-momentum 2-forms are covariantly conserved. Furthermore, since in that case the energy-momentum tensor $ \tau_{ab} $ will be symmetric, i.e $ \tau_{ab} =  \tau_{ba} $, it becomes that the term $ \iota_{n} \tau^{n} = \tau_{nc} \ast ( e^{c} \wedge e^{n} ) = 0 $. In addition, using the expression (\ref{eqn_1_18}) and the symmetry of energy-momentum tensor, one can verify that 
\begin{equation}
\bar{\tau}^{a} = \frac{ 1 }{ 2 } \tau e^{a} + \ast \tau^{a}  \, .
\label{eqn_1_33}
\end{equation} 
Then one can write
\begin{equation}
\epsilon_{abc} \, \tau^{a} \wedge \bar{\tau}^{b} = \frac{ 1 }{ 2 } \tau \epsilon_{abc} \, \tau^{a} \wedge e^{b} + \epsilon_{abc} \, \tau^{a} \wedge \ast \tau^{b} \, .
\label{eqn_1_34}
\end{equation} 
Using the symmetry of energy-momentum tensor, one can also verify that
\begin{equation}
\epsilon_{abc} \, \tau^{a} \wedge e^{b} = 0 
\label{eqn_1_35}
\end{equation}
and 
\begin{equation}
\epsilon_{abc} \, \tau^{a} \wedge \ast \tau^{b} = 0 \, .
\label{eqn_1_36}
\end{equation} 
Then it becomes that the term $ \epsilon_{abc} \, \tau^{a} \wedge \bar{\tau}^{b} = 0 $. For the remaining term $ \epsilon_{abc} \, C^{a} \wedge \bar{\tau}^{b} $, one can write
\begin{equation}
\epsilon_{abc} \, C^{a} \wedge \bar{\tau}^{b} = \frac{ 1 }{ 2 } \tau \epsilon_{abc} \, C^{a} \wedge e^{b} + \epsilon_{abc} \, C^{a} \wedge \ast \tau^{b} \, .
\label{eqn_1_37}
\end{equation}
Furthermore, using the identity $ \iota_{a} C^{a} = 0 $, it can be shown that 
\begin{equation}
\epsilon_{abc} \, C^{a} \wedge e^{b} = C^{a} \wedge \ast ( e_{c} \wedge e_{a} ) = 0 \, .
\label{eqn_1_38}
\end{equation} 
Then
\begin{equation}
\epsilon_{abc} \, C^{a} \wedge \bar{\tau}^{b} = \epsilon_{abc} \, C^{a} \wedge \ast \tau^{b} \, .
\label{eqn_1_39}
\end{equation}
Also noting that 
\begin{equation} 
\epsilon_{abc} \, C^{a} \wedge \ast \tau^{b} = - \epsilon_{abc} \, \tau^{a} \wedge \ast C^{b} \, ,
\label{eqn_1_40}
\end{equation} 
and comparing the coefficients of the terms $ \epsilon_{abc} \, C^{a} \wedge \bar{\tau}^{b} $ and $ \epsilon_{abc} \, \tau^{a} \wedge \ast C^{b} $ on the left and the right hand sides of the expression (\ref{eqn_1_32}) respectively, it can be seen that if the matter Lagrangian is not connection-dependent, then the left and the right hand sides of the consistency relation become identically equal to each other and therefore EGMG field equation with connection-independent matter coupling is third-way consistent without any additional condition. 
 
\section{Minimal and non-minimal matter couplings}

In this section, to illustrate our formalism, we present some samples of minimal and non-minimal matter couplings. We note that minimal bosonic matter couplings are connection-independent in general. On the other hand, non-minimal couplings of bosonic fields to gravity can involve connection-dependent terms in the Lagrangian. For instance, as we will see explicitly in the following, non-minimal couplings of electromagnetic fields to gravity definitely involve connection-dependent terms. As for the fermionic matter couplings such as spinor matter Lagrangians, minimal or non-minimal couplings possess connection-dependancy. We notice that our formalism also allows to consider higher order curvature extensions of EGMG. In fact, for such an extension of EGMG, the independent variation of higher order curvature terms in the Lagrangian with respect to co-frame 1-form $ e^{a} $ and the connection 1-form $ \omega_{ab} $ fields results in the field equations that involve the contractions of curvature terms as well as the covariant derivatives of them. In the resulting field equation of EGMG, such terms contribute to the source term.

\subsection{Minimal matter couplings} 

In this subsection, we provide some samples of minimal matter couplings.

\subsubsection{Connection-independent minimal coupling : }

For the connection-independent minimal couplings, we consider Maxwell-Chern-Simons coupling and scalar field matter coupling with a potential term. Note that in both cases the hyper-momentum forms $ \Omega_{ab} = 0 $ since the matter Lagrangians are connection-independent. Then it becomes that the $ \xi^{ a } = 0 $ and $ \chi^{ a } = 0 $.   

\vspace{0.4cm}

\noindent {\bf i.} Maxwell Chern-Simons Lagrangian 3-form :   

\vspace{0.4cm}

\noindent  As a first sample for the connection-independent matter Lagrangian, we consider the Maxwell Chern-Simons Lagrangian expressed in the form
\begin{equation}
{\cal L}_{EM} [e^{a},  A] = - \frac{ 1 }{2} F \wedge \ast F - \frac{ 1 }{2} m A \wedge F
\label{electromagnetic_lagrangian}
\end{equation}
where $ A $ denotes electromagnetic potential 1-form field while $ F $ describes electromagnetic 2-form field such that $ F = d A $. $ m $ can be identified as the mass parameter associated with Maxwell field. The variation of the Maxwell-Chern-Simons Lagrangian 3-form with respect to co-frame field $ e^{a} $ yields the energy-momentum 2-forms
\begin{equation}
\tau_{c}^{EM} = \frac{ \delta {\cal L}_m [ e^{a} , A ] }{ \delta e^{c} } = \frac{ 1 }{2} ( \iota_{c} F \wedge \ast F - F \wedge \iota_{c} ( \ast F ) ) \, .
\label{maxwell_energy_momentum}
\end{equation}
Furthermore, the trace associated with the energy-momentum 2-form $ \tau_{c} $ can be obtained from
\begin{equation}
\tau  \ast 1 = e^{c} \wedge \tau_{c} = \frac{ 1 }{2} F \wedge \ast F
\end{equation}
where one can use the identity 
\begin{equation}
e^{c} \wedge \iota_{c} \mathit{\Omega} = p \, \mathit{\Omega}
\label{identity}
\end{equation}
for any $p$-form field $ \mathit{\Omega} $. 
Note that by using the expression (\ref{eqn_1_16}), one can easily obtain the term $ \bar{\tau}^{a} $. Then the source term $ \tilde{\tau}_{c} $ of EGMG field equation (\ref{eqn_1_19}) can be calculated from the relation (\ref{source_term}). 

\vspace{0.4cm}

\noindent {\bf ii.} Scalar matter Lagrangian 3-form with a potential term : 

\vspace{0.4cm}
\noindent Our next sample for the connection-independent matter Lagrangian 3-form is the scalar matter Lagrangian that can be expressed in the form
\begin{equation}
{\cal L}_{scalar} [e^{a}, \phi] = - \frac{ 1 }{2} d \phi \wedge \ast d \phi - U(\phi) \ast 1
\label{scalar_field_lagrangian}
\end{equation}   
that includes the potential term $ U(\phi) $ as well. In this case, the variation of scalar matter Lagrangian with respect to co-frame $ e^{a} $ yields energy-momentum 2-forms
\begin{equation}
\tau_{c}^{scalar} = \frac{ \delta {\cal L}_{m} [ e^{a} , \phi] }{\delta e^{c} } = \frac{ 1 }{2} ( \iota_{c} d \phi \wedge \ast d \phi + d \phi \wedge \iota_{c} ( \ast d \phi ) ) - U(\phi) \ast e_{c} \, .
\label{scalar_field_energy_momentum}
\end{equation}
Note that the trace of energy-momentum tensor can be evaluated from the relation
\begin{equation}
\tau \ast 1 = e^{c} \wedge \tau_{c}  = - \frac{ 1 }{2} d \phi \wedge \ast d \phi - 3 U(\phi) \ast 1 
\end{equation}
where again the identity (\ref{identity}) has been used. Then, by using the expression (\ref{scalar_field_energy_momentum}), one can calculate $ \bar{\tau}^{a} $. Then one can calculate the source term of EGMG field equation from the relation (\ref{source_term}).

\subsubsection{Connection-dependent minimal coupling : }

For the connection-dependent matter coupling, we consider spinor-matter coupling described by the Dirac Lagrangian 3-form $ {\cal L}_{D} $ given as
\begin{equation}
{\cal L}_{D} = \frac{i}{2} ( \bar{\psi} \ast e^{a} \gamma_{a} \wedge D \psi - \ast e^{a} \wedge D \bar{\psi} \, \gamma_{a} \, \psi ) - i m \bar{\psi} \psi \ast 1 \, .
\label{dirac_lagrangian}
\end{equation}
One can notice that $ \psi $ denotes the Dirac spinor field while $ \gamma_{a} $ are generators of $ Cl_{1,2} $ Clifford algebra that obeys the relation
\begin{equation}
\{ \gamma_{a} , \gamma_{b} \} = \gamma_{a} \gamma_{b} + \gamma_{b} \gamma_{a} = 2 \eta_{ab} \, .
\end{equation}   
$m$ is the mass parameter associated with spinor field. In addition, we note that the covariant derivative of Dirac spinor field is defined as
\begin{equation}
D \psi = d  \psi + \frac{ 1 }{2} \omega^{cd} \, \sigma_{cd} \psi
\end{equation}
where
\begin{equation}
\sigma_{cd} = \frac{ 1 }{4} [ \gamma_{c} , \gamma_{d} ] = \frac{ 1 }{4} ( \gamma_{c} \gamma_{d} - \gamma_{d} \gamma_{c} ) 
\end{equation}
are generators of algebra of the group $ SO(1,2)$. $ \bar{\psi} $ is Dirac conjugate spinor defined as $ \bar{\psi} = \psi^{\dagger} \gamma_{0} $ whose covariant derivative is given by 
\begin{equation}
D \bar{\psi} =  d \bar{\psi} - \frac{ 1 }{2} \bar{\psi} \sigma_{cd} \, \omega^{cd} \, .
\end{equation}
By making independent variations of the Dirac Lagrangian with respect to co-frame $ e^{a} $ and the connection 1-form $ \omega^{a}\,_{b} $, one can obtain the associated energy-momentum 2-forms $ \tau_{c} $ and the hyper-momentum 2-forms $ \Omega_{ab} $ respectively. They explicitly read
\begin{equation}
\tau_{c} [ e , \omega , \psi ] = \frac{i}{2} ( \ast( e^{a} \wedge e_{c} ) \wedge \bar{\psi} \gamma_{a} D \psi - \ast( e^{a} \wedge e_{c} )  
\wedge D \bar{\psi} \gamma_{a} \psi ) - m ( i \bar{\psi} \psi ) \ast e_{c} \, . 
\label{dirac_energy_momentum}
\end{equation}  
and
\begin{equation}
\Omega^{ab} [ e , \psi ] = - \frac{ 1 }{4} ( i \bar{\psi} \psi ) e^{a} \wedge e^{b} \, . 
\label{dirac_hyper_momentum}
\end{equation}
The trace of the energy momentum tensor can be obtained in the. form
\begin{equation}
\tau \ast 1 = i \left( \ast e^{a} \wedge \bar{\psi}  \gamma_{a} D \psi - \ast e^{a} \wedge D \bar{\psi} \gamma_{a} \psi \right) - 3 m ( i \bar{\psi} \psi ) \ast 1 \, .
\end{equation}
As in the previous matter couplings, one can calculate the term $ \bar{\tau}_{c} $ from the relation (\ref{eqn_1_16}). Furthermore, by using (\ref{eqn_1_14}), the term $ \xi^{a} $ can be calculated as
\begin{equation}
\xi^{a} = - \frac{ 1 }{8} ( i \bar{\psi} \psi ) e^{a} \, .
\label{dirac_xi}
\end{equation} 
Using the expression (\ref{dirac_xi}), one can evaluate the term $ \chi^{a} $ from (\ref{eqn_1_17}) for spinor-matter coupling. It can be obtained in the form
\begin{equation}
\chi^{a} = \frac{ 1 }{8} \ast ( D ( i \bar{\psi} \psi ) \wedge e^{a} ) = \frac{ 1 }{8} \iota^{a} ( \ast D ( i \bar{\psi} \psi ) ) \, .
\label{dirac_chi}
\end{equation}
Then by using (\ref{dirac_xi}), (\ref{dirac_chi}) and (\ref{eqn_1_16}), the source term $ \tilde{\tau}_{c} $ of EGMG field equation with spinor-matter coupling can be obtained from the expression (\ref{source_term}).  

\subsection{Non-minimal matter couplings}

In this part, we consider a sample of non-minimal coupling of bosonic matter in EGMG. One of the couplings that deserve for a particular interest is the non-minimal coupling of electromagnetic fields to gravity. In particular, such a coupling may involve $ R F^{2} $type terms where the Maxwell field non-minimally couples to curvature. In a curved spacetime, the investigation of photon propagation and vacuum polarization effects in quantum electrodynamics (QED) at one-loop level can be described by an effective Lagrangian where $ R F^{2} $-type non-minimal couplings arise. In this case, the Lagrangian with non-minimal coupling can be expressed in the form (also see \cite{baykal} and the references therein.)
\begin{equation}
{\cal L}_{non-minimal} = \nu_{ 1 } R^{ab} \, F_{ab} \wedge \ast F + \nu_{ 2 } P^{a} \wedge \iota_{a} F \wedge \ast F + \nu_{3} R 
F \wedge \ast F 
\label{non_minimal_coupling_1}
\end{equation}   
where $ F_{ab} = \iota_{b} \, \iota_{a} F $ and $ \nu_{ 1 } $, $ \nu_{ 2 } $ and $ \nu_{3} $ are some coupling parameters. 
One can note that such type of couplings can also be obtained from the Kaluza-Klein reduction of quadratic curvature gravity theories \cite{dereli_1}. Also notice that one can express the term $ P^{a} \wedge \iota_{a} F \wedge \ast F $ in the form
\begin{equation}
P^{a} \wedge \iota_{a} F \wedge \ast F = \frac{ 1 }{ 2 } \left( R \, F \wedge \ast F + R^{ab} \, F_{ab} \ast F + R^{ab} \wedge F \wedge \iota_{a} \, \iota_{b}  ( \ast F ) \right) \, .
\label{non_minimal_ricci}
\end{equation} 
Noting that the term 
\begin{equation}
R^{ab} \wedge F \wedge \iota_{a} \, \iota_{b} ( \ast F ) = R^{ab} \wedge F \wedge \ast ( F \wedge e_{b} \wedge e_{a} )  
\label{non_minimal_electromagnetic_identity}
\end{equation}
vanishes in 3-dimensions, the Lagrangian of non-minimal electromagnetic coupling in 3-dimensions can be effectively written in the form
\begin{equation}
{\cal L}_{non-minimal} = \left( \nu_{1} + \frac{\nu_{ 2 }}{ 2 } \right) R^{ab} \, F_{ab} \wedge \ast F + \left( \frac{ \nu_{ 2 }}{ 2 } + \nu_{3} \right) R F \wedge \ast F \, ,
\label{non_minimal_coupling_2}
\end{equation}
that is also confirmed in \cite{unluturk}. Then if one considers the EGMG coupled to Maxwell-Chern-Simons together with non-minimal couplings of electromagnetic fields, the total Lagrangian can be expressed in the form
\begin{equation}
{\cal L} = {\cal L}_{EGMG} + {\cal L}_{EM} + {\cal L}_{non-minimal} 
\label{non_minimal_electromagnetic_egmg}  
\end{equation} 
where the Maxwell-Chern-Simons Lagrangian and the Lagrangian that describes the non-minimal electromagnetic field coupling are given by (\ref{electromagnetic_lagrangian}) and (\ref{non_minimal_coupling_2}) respectively. Then the co-frame variation of the Maxwell-Chern-Simons and the Lagrangian of non-minimal coupling results in the energy-momentum 2-form given by  
\begin{eqnarray}
\tau_{c} & = & \tau_{c}^{EM} + \left( \nu_{1} + \frac{\nu_{ 2 }}{ 2 } \right) \left( \iota_{a} ( F_{bc} R^{ab} \wedge \ast F ) - \iota_{b} ( F_{ac} R^{ab} \wedge \ast F ) \right) \nonumber \\
& & - \frac{ 1 }{ 2 } \left( \nu_{1} + \frac{\nu_{ 2 }}{ 2 } \right) ( \iota_{c} R^{ab} \, F_{ab} \wedge \ast F - F_{ab} \, F \wedge \iota_{c} ( \ast R^{ab} ) ) \nonumber \\
& & -  \frac{ 1 }{ 2 } \left( \nu_{1} + \frac{\nu_{ 2 }}{ 2 } \right) ( \iota_{c} F \, F_{ab} \wedge \ast R^{ab} - R^{ab} \, F_{ab} \wedge \iota_{c} ( \ast F ) ) \nonumber \\
& & + \left( \frac{ \nu_{2} }{ 2 } + \nu_{3} \right) \left( \eta \, \epsilon_{abc} \, R^{ab} - \eta  R \ast e_{c}  - R \, \tau_{c}^{EM} \right)   
\label{non_minimal_energy_momentum}
\end{eqnarray}   
where we define
\begin{equation} 
\eta = \frac{ 1 }{ 2 } F_{ab} \, F^{ab}  
\label{eta}  
\end{equation}
or equivalently $ F \wedge \ast F = \eta \ast 1 $. Note that $ \tau_{c}^{EM} $ is the energy-momentum 2-form associated with Maxwell field given by (\ref{maxwell_energy_momentum}). On the other hand, the connection variation yields the hyper-momentum 2-form expressed in the form
\begin{equation}
\Omega_{ab} = \left( \nu_{ 1 } + \frac{ \nu_{ 2 }}{ 2 } \right) D ( F_{ab} \, \ast F ) + \left( \frac{ \nu_{ 2 }}{ 2 } + \nu_{3} \right) D ( \eta \ast ( e_{a} \wedge e_{b} )  ) \, .
\label{non_minimal_hyper_momentum}
\end{equation}    
Then by using the equation (\ref{eqn_1_16}), one can calculate the term $ \bar{\tau}^{a} $. Furthermore, by using the hyper-momentum expression (\ref{non_minimal_hyper_momentum}), $ \xi^{a} $ and $ \chi^{a} $ can be obtained from the equations (\ref{eqn_1_14}) and (\ref{eqn_1_17}) respectively. Then the source term $ \tilde{\tau}_{c} $ of EGMG field equation with non-minimal electromagnetic coupling can be obtained from the relation (\ref{source_term}). 

\subsection{Higher order curvature extensions of EGMG}

In this part, we illustrate our algebraic method and briefly discuss how to construct the higher order curvature extensions of EGMG. Let us consider that EGMG is coupled with a higher curvature gravity Lagrangian given in the form
\begin{equation}
{\cal L} = {\cal L}_{EGMG} + f ( R , U , V ) \ast 1 \, 
\label{extended_egmg}
\end{equation}  
where the Lagrangian is described by a gravity function $ f ( R , U , V ) $ that depends on the curvature scalar $ R $, the Ricci-squared term $ U $ and the Riemann curvature-squared term $ V $. To be explicit, we have $ U \ast 1 = P_{a} \wedge \ast P^{a} $ and $ V \ast 1 = 2 R_{ab} \wedge \ast R^{ab} $ respectively. Note that the terms obtained from the independent variations of the higher curvature gravity Lagrangian with respect to co-frame and the connection fields contribute to the source term of extended EGMG field equation. Now the independent co-frame variation of the higher curvature gravity Lagrangian yields the energy-momentum 2-form given as 
\begin{eqnarray}
\tau_{c} & = & \left( \frac{ \partial f }{ \partial R} \right)  \, \epsilon_{abc} R^{ab} - \left( \frac{ \partial f}{ \partial R} \right) \, R \ast e_{c} - 2 \left( \frac{ \partial f}{ \partial U } \right)  \, \iota_{b} ( \iota_{c} R^{ba} \wedge \ast P_{a} ) \nonumber \\ 
 & & - \left( \frac{ \partial f}{ \partial U } \right) \, ( \iota_{c} P_{a} \wedge \ast P^{a} + P_{a} \wedge \iota_{c} ( \ast P^{a} ) 
 - \left( \frac{ \partial f }{ \partial U } \right) \, U \ast e_{c} \nonumber \\
 & & - 2 \left( \frac{ \partial f}{ \partial V} \right) ( \iota_{c} R_{ab} \wedge \ast R^{ab} - R_{ab} \wedge \iota_{c} ( \ast R^{ab} ) - \left( \frac{ \partial f}{ \partial V} \right) \, V \ast e_{c} \nonumber \\ 
 & & + f \ast e_{c} \, .
\label{extended_curvature_energy_momentum}     
\end{eqnarray}       
On the other hand, the independent connection variation yields the hyper-momentum 2 -form
\begin{eqnarray}
\Omega_{ab} & = & D \left( \left( \frac{ \partial f}{\partial R } \right) \ast ( e_{a} \wedge e_{b} ) \right) + D \left( \left( \frac{ \partial f}{ \partial U } \right) \ast ( P_{a} \wedge e_{b} - P_{b} \wedge e_{a} ) \right) \nonumber \\
& & + 4 D \left( \left( \frac{ \partial f}{ \partial V} \right) \ast R_{ab} \right) \, .
\label{extended_curvature_hyper_momentum}
\end{eqnarray}
As in the non-minimal electromagnetic coupling, one can use the equation (\ref{eqn_1_16}) and calculate the term $ \bar{\tau}^{a} $. Furthermore, by using the hyper-momentum expression (\ref{extended_curvature_hyper_momentum}), $ \xi^{a} $ and $ \chi^{a} $ can be evaluated from the equations (\ref{eqn_1_14}) and (\ref{eqn_1_17}) respectively. Then the source term $ \tilde{\tau}_{c} $ of EGMG field equation extended with higher order curvature couplings can be obtained from the relation (\ref{source_term}). In particular, one can consider the quadratic curvature gravity extension of EGMG where one can express the quadratic curvature gravity Lagrangian in the form
\begin{equation}
{\cal L}_{QCG} = \frac{ 1 }{ 2 } \alpha_{ 1 } R^{ 2 } \ast 1 + \frac{ 1 }{ 2 } \alpha_{ 2 } P_{a} \wedge \ast P^{a} + \frac{ 1 }{ 2 } \alpha_{3} R_{ab} \wedge \ast R^{ab} \, .
\label{quadratic_curvature}
\end{equation}
Note that the case for which the coupling parameters become $ \alpha_{ 1 } = - \frac{3}{8} \alpha_{2} $ and $ \alpha_{3} = 0 $ corresponds to the New Massive Gravity Lagrangian. For the quadratic curvature gravity Lagrangian (\ref{quadratic_curvature}), the gravity function $ f ( R , U , V ) $ becomes
\begin{equation}
f ( R , U , V ) = \frac{ 1 }{ 2 }  \alpha_{ 1 }  R^{ 2 } + \frac{ 1 }{ 2 }  \alpha_{ 2 } U + \frac{ 1 }{4}  \alpha_{3} V  
\label{gravity_function}
\end{equation}   
Then the energy-momentum and hyper-momentum 2-forms become
\begin{eqnarray}
\tau_{c} & = &  \alpha_{ 1 } \epsilon_{abc} \, R \, R^{ab} - \frac{ 1 }{ 2 } \alpha_{ 1 }  R^{ 2 } \ast e_{c} - \alpha_{ 2 } \iota_{b} ( \iota_{c} R^{ba} \wedge \ast P_{a} ) \nonumber \\
& & - \frac{ 1 }{ 2 } \alpha_{ 2 } ( \iota_{c} P_{a} \wedge \ast P^{a} + P_{a} \wedge \iota_{c} ( \ast P^{a} ) ) \nonumber \\
& & - \frac{ 1 }{ 2 } \alpha_{3} ( \iota_{c} R_{ab} \wedge \ast R^{ab} - R_{ab} \wedge \iota_{c} ( \ast R^{ab} ) ) 
\label{quadratic_curvature_energy_momentum}
\end{eqnarray}
and
\begin{equation}
\Omega_{ab} = \alpha_{ 1 } D ( R \, \ast ( e_{a} \wedge e_{b} ) ) + \frac{ 1 }{ 2 } \alpha_{ 2 } D ( \ast ( P_{a} \wedge e_{b} - P_{b} \wedge e_{a} ) ) + \alpha_{3} D ( \ast R_{ab} ) 
\label{quadratic_curvature_hyper_momentum} 
\end{equation} 
respectively. Then by using the equation (\ref{eqn_1_16}), one can calculate the term $ \bar{\tau}^{a} $ for the quadratic curvature gravity. In addition, by using (\ref{quadratic_curvature_hyper_momentum}), $ \xi_{a} $ and $ \chi_{a} $ can be obtained from the equations (\ref{eqn_1_14}) and (\ref{eqn_1_17}) respectively. Then using the expression (\ref{source_term}), one can construct the source 2-form of EGMG field equation extended with quadratic curvature terms. We note that a detailed investigation of such an extension of EGMG certainly deserves for a future work. Especially, it would be interesting to investigate the unitarity properties and the propagating modes of the theory within our formalism.    

\section{ Conclusion } 

In this work, using exterior algebra notation, we examine the matter coupling in EGMG considering that the matter Lagrangian is both co-frame and connection dependent. By making independent variations of matter-coupled EGMG Lagrangian with respect to the co-frame, the connection and the auxiliary fields, we derive the field equations from a variational principle. Then by algebraically solving the auxiliary 1-form fields in terms of gravitational field variables, energy-momentum and hyper-momentum 2-forms, we derive and construct source 2-forms of matter-coupled EGMG field equation. We illustrate that the source 2-form definitely involves quadratic terms in energy-momentum and hyper-momentum 2-forms as well as their covariant derivatives. Next, concentrating on the consistency of matter-coupled EGMG field equation, we derive the consistency relation satisfied by the source 2-form. Evaluating the left and the right hand sides of the consistency relation, we see that in general both sides of the consistency relation are not identically equal which implies that an additional relation should hold between the energy-momentum and the hyper-momentum 2-forms in order that EGMG field equation be third way consistent. In the next part, we provide some sample matter Lagrangians to illustrate our formalism. In fact, one can see that our formalism allows both minimal and non-minimal matter couplings with gravitational fields. As a particular example, we illustrate the non-minimal coupling of electromagnetic fields to gravity where the non-minimal electromagnetic coupling can involve the curvature couplings with Maxwell field. Then it can be seen that the Lagrangian of such a non-minimal coupling depends on both co-frame and the connection fields together with the Maxwell field. We further note that by evaluating energy-momentum and hyper-momentum 2-forms, the auxiliary fields can be algebraically solved even for the case of non-minimal coupling. We consider that a detailed investigation of EGMG with non-minimal electromagnetic coupling deserves for a future work.             

Apart from non-minimal matter couplings, it can be noticed our formalism also offers to construct higher order curvature extensions of EGMG. It can be seen that independent co-frame and connection variations of EGMG extended with higher order curvature terms yield energy-momentum and hyper-momentum 2-form expressions that involve the curvature terms together with the contractions and the covariant derivatives of curvature terms. Then as for the case of non-minimal coupling, the auxiliary fields can again be algebraically obtained and as a result the effective source 2-form of EGMG field equation extended with higher order curvature terms can be derived. In particular, we illustrate the case where EGMG is extended with quadratic curvature terms. We again note that a detailed investigation of EGMG extended with higher order curvature terms deserve for a future study. In fact, it would be interesting to examine BTZ and warped AdS solutions and calculate conserved charges.             

Furthermore, one can investigate the consistent matter coupling in some other exotic massive gravity models obtained from the truncation of auxiliary fields \cite{afshar_2}. In particular, one can study the matter coupling in the so-called extended exotic massive gravity constructed in \cite{afshar_2} by the method of truncation.Note that the auxiliary field $ \lambda^{a} $ introduced in EGMG can also be identified as the Lagrange multiplier 1-form field that enforces the zero-torsion constraint in the gravity theory. In fact, one can release this constraint and examine the matter coupling in new exotic type and MMG-like gravity models involving non-vanishing torsion where such type of gravity models have been recently presented in \cite{ozkan_4} and very recently introduced in \cite{ozkan_5} respectively, using a particular truncation of infinite dimensional Lie algebra \cite{ozkan_3}.  For all such type of gravity models, it would be interesting to construct the source 2-forms of matter-coupled field equation and to obtain the consistency relation. Apart from matter couplings, it would also be of great interest to construct the Weyl-covariant formulation of exotic gravity models in non-metric compatible 3-dimensional spacetimes \cite{dereli_2}.  

Another interesting future work is the construction of supersymmetric version of EGMG. In fact, we consider that the methods employed in this work are proven to be useful and can lead to a consistent Exotic General Massive Supergravity theory where a supersymmetric version of MMG has recently been presented in \cite{deger_5}.    

For other future directions and final remarks, one can also consider to obtain some exact solutions such as pp-waves, warped AdS black holes etc., of matter-coupled EGMG. In particular, one can obtain warped AdS black hole solutions of EGMG coupled to Maxwell-Chern-Simons where such type of a black hole solution has been previously constructed in \cite{nam} for MMG coupled to Maxwell-Chern-Simons. In addition, it would be interesting to get pp-wave solutions of EGMG with Maxwell-Chern-Simons coupling where a pp-wave solution has been previously given in \cite{dereli_3} for TMG and a pp-wave solution of MMG coupled to Maxwell-Chern-Simons has been recently introduced in \cite{cebeci_2}. Apart from exact solutions of matter-coupled EGMG, one can consider fermionic couplings in EGMG and realize a detailed investigation of EGMG with spinor-matter coupling. These would again be the subjects of future research.

%\section*{Acknowledgments }  

%\appendix

%\section{The method of solving the auxiliary 1-form fields $\rho^{a}$ and $ \lambda^{a}$ algebraically}

\end{document}